\documentclass[aps,12pt,showpacs,showkeys,tightenlines]{revtex4}

\usepackage{amsmath}
\usepackage{amssymb,amsfonts}
\usepackage{bm}

\newcommand{\etal}{\textit{et~al.}}
\newcommand{\ie}{\textit{i.e.}}

\newcommand{\mathnotation}[2]{\newcommand{#1}{#2}}

\renewcommand{\l}{\left}			
\renewcommand{\r}{\right}			
\mathnotation{\pd}{\partial}			
\mathnotation{\ee}{{\mathrm e}}			
\mathnotation{\grad}{{\boldsymbol{\nabla}}}	
\mathnotation{\lapl}{\nabla^2}			
\mathnotation{\curl}{\grad\times}		
\renewcommand{\div}{\grad\cdot}			
\mathnotation{\ldef}{\mathrel{\raisebox{.069ex}{:}\!\!=}}
\mathnotation{\rdef}{\mathrel{=\!\!\raisebox{.069ex}{:}}}
\mathnotation{\half}{\tfrac{1}{2}}		
\mathnotation{\levicivita}{\varepsilon}		
\newcommand{\Order}[1]{\mathrm{O}\!\l(#1\r)}	
\renewcommand{\time}{t}				
\mathnotation{\lyapexp}{\lambda}		
\mathnotation{\x}{x}				
\mathnotation{\xv}{{\bm{\x}}}		
\mathnotation{\vel}{{\bm{v}}}		
\mathnotation{\velc}{v}				
\mathnotation{\sdim}{n}				
\mathnotation{\lac}{a}				
\mathnotation{\lav}{{\bm{\lac}}}		
\mathnotation{\metric}{g}			
\mathnotation{\detmetric}{g}			
\mathnotation{\geigen}{\Lambda}			
\mathnotation{\udir}{{\mathrm{u}}}		
\mathnotation{\mdir}{{\mathrm{m}}}		
\mathnotation{\sdir}{{\mathrm{s}}}		
\mathnotation{\udirv}{{\mathbf{u}}}
\mathnotation{\mdirv}{{\mathbf{m}}}
\mathnotation{\sdirv}{{\mathbf{s}}}
\mathnotation{\udiru}{\hat{\udir}}
\mathnotation{\mdiru}{\hat{\mdir}}
\mathnotation{\sdiru}{\hat{\sdir}}
\mathnotation{\udiruv}{\hat{\udirv}}
\mathnotation{\mdiruv}{\hat{\mdirv}}
\mathnotation{\sdiruv}{\hat{\sdirv}}
\mathnotation{\ediru}{\hat{\mathrm{e}}}
\mathnotation{\ediruv}{\hat{\mathbf{e}}}
\mathnotation{\gradlc}{\grad_0}			
\mathnotation{\curllc}{\gradlc\times}		
\mathnotation{\divlc}{\gradlc\cdot}		
\mathnotation{\Jacm}{M}				

\mathnotation{\B}{B}				
\mathnotation{\Bv}{\bm{B}}
\mathnotation{\A}{A}				
\mathnotation{\Av}{\bm{A}}
\mathnotation{\current}{j}			
\mathnotation{\currentv}{\bm{\current}}
\mathnotation{\currentp}{\current_\parallel}
\mathnotation{\helic}{{\mathcal{H}}}		
\mathnotation{\bb}{b}
\mathnotation{\bbv}{\bm{\bb}}
\mathnotation{\resist}{\eta}
\mathnotation{\EnerB}{E_{\B}}			
\mathnotation{\permbfs}{\mu_0}
\mathnotation{\dissts}{\tau_{\mathrm{d}}}
\mathnotation{\maxnug}{\mathcal{K}}
\mathnotation{\maxnugg}{\mathcal{J}}

\begin{document}

\title{The Onset of Dissipation in the Kinematic Dynamo}
\date{20 September 2002}
\author{Jean-Luc Thiffeault}
\email{jeanluc@mailaps.org}
\author{Allen H. Boozer}
\affiliation{Department of Applied Physics and Applied
Mathematics, Columbia University, New York, NY 10027}

\begin{abstract}
The kinematic regime of the magnetic dynamo neglects the backreaction of the
magnetic field on the flow.  For small magnetic diffusivity, in the early
stage of evolution, there is an ideal phase where dissipative effects can also
be neglected.  We estimate the magnitude of the energy dissipation term (Ohmic
heating), taking into account differential constraints on chaotic flows.  We
find that the period of ideal evolution is roughly doubled over an estimate
without constraints.  The helicity generation terms are exponentially smaller
than the energy dissipation, so that large quantities of energy are dissipated
before any helicity can be created.  Helicity flow is exponentially larger
than net helicity generation.  The constraints also lead to the existence of a
singular initial condition for the magnetic field for which sizable amounts of
helicity can potentially be created.
\end{abstract}

\maketitle

\section{Introduction}
\label{eq:intro}

For many three-dimensional flows an embedded magnetic field will grow
exponentially; this phenomenon is referred to as a dynamo~\cite{STF}.  The
kinematic dynamo problem consists of studying the induction equation for the
field on the assumption that it does not react back on the flow (the Lorentz
force is neglected).  This is justified as long as the field is small enough
compared to inertial forces.  A good estimate of the domain of validity of the
kinematic assumption is thus of paramount importance.

It was recently pointed out by Schekochihin \etal~\cite{Schekochihin2002} that
the details of how the nonlinear Lorentz force term appears in the equations
of motion are important.  The Lorentz tension force only involves the gradient
of the magnetic field along itself (the parallel gradient), and in smooth
chaotic flows or turbulent flows at large Prandtl number that gradient is much
smaller than its perpendicular counterpart.  Thus, the onset of the reaction
force is delayed as compared with an estimate based on the scale and magnitude
of the magnetic field.

In this paper we investigate the domain of validity of the \emph{ideal}
evolution assumption, that the field is passively advected by the plasma with
resistive effects being negligible.  Because the resistivity (magnetic
diffusivity) is assumed very small in dimensionless terms (\ie, the magnetic
Reynolds number is very large, typically~$10^8$--$10^{15}$; for example, it is
of order~$10^{12}$ in the solar corona), there is a sizable period of
evolution where the scale of the magnetic field is such that Ohmic dissipation
is unimportant.  In a flow with chaotic Lagrangian trajectories the
exponential stretching of fluid elements leads to growth of the amplitude of
the magnetic field, so that the magnetic energy grows very rapidly during the
period of ideal evolution.  But in incompressible flows there is a
corresponding exponential decrease via folding of the scale of magnetic field
variations.  These small scale variations in the field cause the Ohmic
dissipation term to become large, and the evolution is no longer ideal.  A
convenient method of estimating the importance of the nonideal (resistive)
term is to compare the rate of growth of magnetic energy to that of the power
dissipated by Ohmic heating.  This was done in
Refs.~\cite{Boozer1992,Tang2000} where it was found that, because the power
dissipation involves extra gradients of the magnetic field, the ratio of
energy over power (the dissipation time scale) decreases exponentially in time
at a rate~$-4\lyapexp$, where~$\lyapexp$ is a typical value of the Lyapunov
exponent of the flow---the mean rate of exponential stretching of fluid
elements.

In the present paper we will show that these estimates are overly pessimistic.
Because of the existence of \emph{differential constraints} on the
characteristic directions and amplitudes of stretching, the dissipation time
scale decreases exponentially at a typical rate~$-2\lyapexp$, implying that
the period of ideal evolution is twice as long as previously thought.  These
differential constraints originated in the work of Tang and
Boozer\cite{Tang1996} on the advection-diffusion equation~, and Thiffeault and
Boozer~\cite{Thiffeault2001} found a new constraint valid in three dimensions
that we apply here to the kinematic dynamo.  Methods from differential
geometry were used to show that the derivatives along the characteristic
directions of expansion of the stretching rates cannot be completely
independent and have to satisfy constraints.  The theory of differential
constraints was recently expanded and put on a more rigorous footing by
Thiffeault~\cite{Thiffeault2002}.  The constraints have been tested
numerically to a high degree of precision for various maps and
flows~\cite{Tang1996,Thiffeault2001,Thiffeault2002}.

After reviewing aspects of stretching in chaotic flows in
Section~\ref{sec:stretching}, we calculate the energy dissipation time scale
in the manner of Boozer~\cite{Boozer1992} in Section~\ref{sec:ejtraj} and
introduce the necessary notation along the way.  The effect of differential
constraints on the dissipation time scale is derived in
Section~\ref{sec:constr}.  In Section~\ref{sec:singinit} we investigate an
intriguing consequence of the differential constraint, namely that the
exists a \emph{singular initial condition} for which the energy dissipation
time scale does not decrease exponentially; we speculate on the physical
meaning of such an initial condition.  Section~\ref{sec:helicity} is devoted
to deriving estimates of the amount of helicity created during the period of
ideal growth, both with and without constraints and for the singular initial
condition.  Finally, in Section~\ref{sec:discussion} we offer some general
comments on the results of the paper and discuss future work.

\section{Stretching in Chaotic Flows}
\label{sec:stretching}

In this section we review the properties of stretching in chaotic flows, as
embodied by a metric tensor in Lagrangian coordinates that measures the total
deformation experienced by a fluid element as it is advected and stretched by
the flow.

For a given smooth velocity field~$\vel(\xv,\time)$, the trajectory of a fluid
element is the solution to
\begin{equation}
	\dot\xv(\lav,\time)
	= \vel(\xv(\lav,\time),\time),
	\qquad \xv(\lav,0) = \lav,
\end{equation}
where~$\xv(\lav,\time)$ is the position at time~$\time$ of a fluid element
that started at~$\lav$ at~$\time=0$.  The initial condition labels~$\lav$ are
called Lagrangian coordinates, and the~$\xv$ are the Eulerian coordinates.
(The Eulerian coordinates are ``fixed'' in space and can be regarded as
ordinary Cartesian coordinates.)  The function~\hbox{$\xv = \xv(\lav,\time)$}
is thus the transformation from Lagrangian ($\lav$) to Eulerian ($\xv$)
coordinates.  For a chaotic flow, this transformation gets extremely
complicated as time evolves.

The \emph{Jacobian matrix} of the transformation $\xv(\lav,\time)$
is~\hbox{${\Jacm^i}_q \ldef \pd\x^i/\pd\lac^q$}.  The Jacobian matrix is a
precise record of how an initially spherical fluid element is rotated and
stretched into an ellipsoid by the flow.  We are interested in describing
stretching and its direction in the reference frame of a fluid element
(Lagrangian frame), not the absolute rotation in Eulerian space, so we
construct the \emph{metric tensor} or \emph{Cauchy--Green strain
tensor}~\cite{Ottino},~\hbox{$\metric_{pq} \ldef
\sum_{i=1}^{3}\,{\Jacm^i}_p\,{\Jacm^i}_q$}, which contains only the
information on the stretching of fluid elements (this can be seen, for
example, by performing a singular value decomposition
of~$\Jacm$~\cite{Greene1987,Geist1990,Thiffeault2002}).  The metric tensor is
a symmetric, positive-definite matrix, so it can be diagonalized with
orthogonal eigenvectors $\{\udiruv,\mdiruv,\sdiruv\}$, and corresponding real,
positive eigenvalues~$\{\geigen_\udir^2,\geigen_\mdir^2,\geigen_\sdir^2\}$; we
can thus write
\begin{equation}
	\metric_{pq} = \geigen_{\udir}^2\,\udiru_p\,\udiru_q
		+ \geigen_{\mdir}^2\,\mdiru_p\,\mdiru_q
		+ \geigen_{\sdir}^2\,\sdiru_p\,\sdiru_q.
	\label{eq:metricdiag}
\end{equation}
The~$\geigen$'s and~$\{\udiruv,\mdiruv,\sdiruv\}$ are functions of~$\lav$
and~$\time$ (but see below).  The~$\geigen$'s are called \emph{coefficients of
expansion}.  Without loss of generality, we order them such
that~$\geigen_{\udir} > \geigen_{\mdir} > \geigen_{\sdir}$ (we assume
nondegeneracy of the eigenvalues).  The label~``$\udir$'' indicates an
unstable direction: after some time, we have~\hbox{$\geigen_\udir \gg 1$},
growing exponentially for large time.  Such a direction exists if the flow has
chaotic trajectories, which we assume to be the case.  The label~``$\sdir$''
indicates a stable direction: after some time, we have~\hbox{$\geigen_\sdir
\ll 1$}, decreasing exponentially for long times.  The intermediate direction,
denoted by~``$\mdir$'', can grow or decrease exponentially at a rate somewhere
between that of~$\geigen_\udir$ and~$\geigen_\sdir$; for steady flows (\ie,
when $\vel$ is independent of~$\time$),~$\geigen_\mdir$ has only algebraic
(nonexponential) behavior~\cite{Eckmann1985}.  The determinant of the metric
is denoted by~$\detmetric$; if~$\vel$ is incompressible ($\div\vel=0$), we
have~\hbox{$\sqrt{\detmetric} = \geigen_\udir\,\geigen_\mdir\,\geigen_\sdir =
1$}.

The eigenvectors $\{\udiruv,\mdiruv,\sdiruv\}$ are called characteristic
directions of stretching.  It is well-known that these characteristic
directions converge exponentially in to time-asymptotic values that depend
only on the initial condition~$\lav$ (and the initial time for unsteady
flows)~\cite{Goldhirsch1987,Thiffeault2002}.  We assume that these directions
(on the trajectory of interest) have converged.  It is also well-known that
the Lyapunov exponents converge far slower than the characteristic direction.
We do \emph{not} assume that the finite-time Lyapunov
exponents~$\lyapexp_\sigma(\lav,\time)$, defined by
\begin{equation}
  \lyapexp_\mu \ldef \frac{1}{\time}\,
  \log\geigen_\mu\,,
  \label{eq:lyapexpdef}
\end{equation}
have converged or even that their limit as~$\time\rightarrow\infty$ exists; we
merely suppose that~$\time$ is sufficiently large to have~$\geigen_\udir\gg1$
and~$\geigen_\sdir\ll1$.  This difference in magnitude between the unstable
and stable directions is the basis for the approximations used in this paper.

\section{Energy and Current along Trajectories}
\label{sec:ejtraj}

Equipped with the metric tensor in diagonal form~\eqref{eq:metricdiag}, we can
now compute the evolution of physical quantities, such as the energy and
current, along fluid trajectories.  This was investigated in
Refs.~\cite{Boozer1992,Tang2000}, where the relative growth rates of energy,
parallel current, etc. were derived.  In the present section we revisit these
results; in Section~\ref{sec:constr} we will show that they must be modified
to account for \emph{differential constraints}.

The evolution of a magnetic field in resistive MHD is governed by the
induction equation,
\begin{equation}
	\frac{\pd\Bv}{\pd\time} = \curl(\vel\times\Bv)
		+ \frac{\resist}{\permbfs}\lapl\Bv
	\label{eq:induction}
\end{equation}
where $\Bv$ is the magnetic field, $\resist$ is the resistivity, and
$\permbfs$ is the permeability of free space.  With the help of the chain rule
and the metric tensor, we can transform the magnetic induction
equation~\eqref{eq:induction} from Eulerian coordinates, $\xv$, to Lagrangian
coordinates, $\lav$,
\begin{equation}
	{\l.\frac{\pd}{\pd\time}\r|}_{\lav}\bb^r(\lav,\time)
	= \sum_{p,q=1}^3\frac{\resist}{\permbfs}\,\frac{\pd}{\pd\lac^p}
	\l[\metric^{pq}(\lav,\time)\,\frac{\pd}
		{\pd\lac^q}\bb^r(\lav,\time)\r]
	\label{eq:inductionlagrc}
\end{equation}
where $\bb^r/\sqrt{\detmetric} \ldef \sum_i{{(\Jacm^{-1})}^r}_i\,\B^i$ is the
magnetic field in the Lagrangian frame, and $\metric^{pq} \ldef
({\metric^{-1}})^{pq}$.  The~$|_\lav$ subscript on~$\pd/\pd\time$ is a
reminder that the time derivative is taken holding~$\lav$ constant, as opposed
to the derivative in~\eqref{eq:induction} which has~$\xv$ fixed.

Equation~\eqref{eq:inductionlagrc} is simply a diffusion equation with
anisotropic diffusivity tensor~$(\resist/\permbfs)\metric^{pq}$.  By
construction, the velocity $\vel$ has dropped out of the equation entirely.
When $\resist=0$, we have the well-known Cauchy solution~\hbox{$\bbv =
\bbv(\lav)$}, independent of time, so that~$\bbv$ is the magnetic field
at~$\time=0$.  This is the classic result that in ideal MHD the magnetic field
is frozen into the plasma.

In the absence of diffusivity, the magnetic field is amplified in a chaotic
flow, as can be seen by considering the magnetic energy~\hbox{$\EnerB \ldef
(1/2\permbfs)\B^2 = \metric_{pq}\,\bb^p\,\bb^q/2\permbfs\detmetric$}.  From
Eq.~\eqref{eq:metricdiag}, the dominant part of~$\B^2$
is~\hbox{$\geigen_\udir^2\,(\bbv\cdot\udiruv)^2/\detmetric$}, which in a
chaotic flow grows exponentially with time.

In Lagrangian coordinates, the induced
current~\hbox{$\currentv=\curl\Bv/\permbfs$} is~\cite{Boozer1992}
\begin{equation}
	\permbfs\,\current^r
	= \sum_{p,q,q'}\frac{\levicivita^{rpq}}{\sqrt\detmetric}
	\frac{\pd}{\pd\lac^p}
	\l(\metric_{qq'}\,\frac{\bb^{q'}}{\sqrt\detmetric}
	\r),
\end{equation}
which after using Eq.~\eqref{eq:metricdiag} and projecting along
the~$\{\udiruv,\mdiruv,\sdiruv\}$ basis becomes
\begin{align}
\permbfs\,\current_\udir &= \phantom{-}\geigen_\udir^2\,\bb_\udir
		\,\udiruv\cdot\curllc\udiruv
	- \geigen_\mdir^2\,\bb_\mdir\,\sdiruv\cdot\l[
		\gradlc\log\l(\geigen_\mdir^2|\bb_\mdir|\r) -
		(\mdiruv\cdot\gradlc)\mdiruv\r]
	+ \Order{\geigen_\sdir^2},
	\label{eq:judir}\\
\permbfs\,\current_\mdir &= \phantom{-}
	\geigen_\udir^2\,\bb_\udir\,\sdiruv\cdot\l[
		\gradlc\log\l(\geigen_\udir^2|\bb_\udir|\r) -
		(\udiruv\cdot\gradlc)\udiruv\r]
	+ \Order{\geigen_\mdir^2},
	\label{eq:jmdir}\\
\permbfs\,\current_\sdir &=
	-\geigen_\udir^2\,\bb_\udir\,\mdiruv\cdot\l[
		\gradlc\log\l(\geigen_\udir^2|\bb_\udir|\r) -
		(\udiruv\cdot\gradlc)\udiruv\r]
	+ \Order{\geigen_\mdir^2},
	\label{eq:jsdir}
\end{align}
where~$\current_{\udir} \ldef (\sqrt{\detmetric}\,\udiruv\cdot\currentv$),
$\bb_\udir \ldef (\udiruv\cdot\bbv/\sqrt{\detmetric})$, etc.  The subscript
on~$\gradlc$ denotes differentiation with respect to the Lagrangian
coordinates,~$\lav$.  In Refs.~\cite{Boozer1992,Tang2000} it was concluded
that the dominant term in~\hbox{$\detmetric\,\current^2 =
\geigen_\udir^2\,\current_\udir^2 + \geigen_\mdir^2\,\current_\mdir^2 +
\geigen_\sdir^2\,\current_\sdir^2$} arises
from~$\geigen_\udir^2\,\current_\udir^2$, so that overall the magnitude of the
current grows as~$\geigen_\udir^6/\detmetric^2$.  This is an extremely rapid
growth: following Ref.~\cite{Boozer1992}, we define the dissipation time
scale,
\begin{equation}
	\dissts \ldef \EnerB/\resist\current^2,
	\label{eq:disstsdef}
\end{equation}
the ratio of the magnetic energy to the power dissipated in the plasma through
Ohmic heating, and find that~$\dissts\sim\detmetric\geigen_\udir^{-4}$.  Of
course, because the Cauchy solution for the magnetic field is only valid for
ideal evolution---before the dissipation comes into play---the time
scale~$\dissts$ must be regarded as an indicator of the accumulation of small
scale gradients.  When~$\dissts$, which is initially large because of the
small resistivity, becomes of the order of typical macroscopic time scales,
the ideal evolution approximation becomes invalid.  That~$\dissts$ evolves
as~$\detmetric\geigen_\udir^{-4}$ indicates that the ideal evolution must be
abandoned rather quickly, and that relatively little magnetic energy has
accumulated when this time is reached.  We will see in the following section
that the situation is not as hopeless as it appears: the actual domain of
validity of ideal evolution is considerably greater than the calculation of
this section indicates.

\section{The Effect of Differential Constraints}
\label{sec:constr}

The conclusions of Section~\ref{sec:ejtraj} regarding the decay of~$\dissts$
was based on the assumption that~$\udiruv\cdot\curllc\udiruv$
in~\eqref{eq:judir} was a nonexponential function.  In
Refs.~\cite{Thiffeault2001,Thiffeault2002} it was shown that this is not the
case: in three dimensions we must have
\begin{equation}
	\udiruv\cdot\curllc\udiruv \sim
	\geigen_\udir^{-2}\max\l(\sqrt{\detmetric}\,,\geigen_\mdir^2\r)
	\longrightarrow 0,
	\label{eq:constr1}
\end{equation}
as long as the flow is chaotic and there exists a contracting
direction,~$\sdiruv$.  Equation~\eqref{eq:constr1} is one example of a
\emph{differential constraint} in chaotic flows; it is realized with
exponential accuracy in time.  Because it will recur often, we define the
factor
\begin{equation}
	\maxnug \ldef \max\l(\sqrt{\detmetric}\,,\geigen_\mdir^2\r);
	\label{eq:maxnug}
\end{equation}
For a steady incompressible flow, the growth or decay of~$\maxnug$ is
algebraic, not exponential, so in that case we can neglect~$\maxnug$ compared
to exponential factors to get an estimate of growth rates.

The constraint~\eqref{eq:constr1} and others [see below,
Eqs.~\eqref{eq:constr2}--\eqref{eq:constr3}] were derived in
Ref.~\cite{Thiffeault2001} using the condition that the Riemannian curvature
of the metric tensor~$\metric_{pq}$ vanishes.  In Ref.~\cite{Thiffeault2002}
the constraints were derived under more general assumptions, and they were
shown to be present in arbitrary dimensions; their convergence rate was also
obtained.  The reason why~\eqref{eq:constr1} holds is not readily apparent;
it is due to the overconstrained nature of the Lagrangian derivatives of the
characteristic directions and coefficients of
expansions~\cite{Thiffeault2002}.  All the constraints used in this paper have
been verified numerically to a high degree of
precision~\cite{Tang1996,Thiffeault2001,Thiffeault2002} for a variety of maps
and flows.

In three dimensions, there are two more constraints in addition
to~\eqref{eq:constr1}~\cite{Thiffeault2001,Thiffeault2002},
\begin{align}
	\udiruv\cdot(\udiruv\cdot\gradlc)\sdiruv + \sdiruv\cdot\gradlc
		\log\geigen_\udir &\sim \geigen_\sdir \longrightarrow 0,
	\label{eq:constr2}\\
	\mdiruv\cdot(\mdiruv\cdot\gradlc)\sdiruv + \sdiruv\cdot\gradlc
		\log\geigen_\mdir &\sim
		\max(\geigen_\sdir\,,\geigen_\sdir^2/\geigen_\mdir^2)
		\longrightarrow 0.
	\label{eq:constr3}
\end{align}
If we add~\eqref{eq:constr2} and~\eqref{eq:constr3}, we find after some
manipulation
\begin{equation}
	\frac{1}{\sqrt{\detmetric}}\divlc\l(\sqrt{\detmetric}\,\sdiruv\r)
	- \sdiruv\cdot\gradlc\log\geigen_\sdir \sim
	\max(\geigen_\sdir\,,\geigen_\sdir/\geigen_\mdir) \longrightarrow 0;
	\label{eq:divsconstr}
\end{equation}
This is a compressible ($\sqrt{\detmetric}\ne1$) version of the constraint
used in Ref.~\cite{Tang2000}, originally derived in two dimensions in
Ref.~\cite{Tang1996}.  In two dimensions it is the only independent
constraint.

In addition, if it happens that the~$\mdiruv$ direction (the middle
eigenvalue) is a contracting direction (\ie, $\geigen_\mdir$ decreases
exponentially for large times), then there is a fourth constraint,
\begin{equation}
	\udiruv\cdot(\udiruv\cdot\gradlc)\mdiruv + \mdiruv\cdot\gradlc
		\log\geigen_\udir \sim \geigen_\mdir \longrightarrow 0,
	\quad\text{for\ }\geigen_\mdir\ll 1.
	\label{eq:constr4}
\end{equation}
Thus there are three constraints,~\eqref{eq:constr1},~\eqref{eq:constr2},
and~\eqref{eq:constr3}, if~$\geigen_\mdir$ is a neutral or stretching
direction, and four if~$\geigen_\mdir$ is a contracting direction.  For a
generic three-dimensional chaotic flow satisfying the assumptions herein (that
is, at least one stretching and one contracting direction), there can be no
further differential constraints that involve only first derivatives of the
characteristic directions and coefficients of expansion~\cite{Thiffeault2002}.
It is however possible that flows with special symmetries admit additional
relationships (algebraic, differential, or integro-differential) among the
characteristic directions and coefficients of expansion.  One example is that
for incompressible flows the coefficients of expansion obey the algebraic
relationship~\hbox{$\geigen_\udir\,\geigen_\mdir\,\geigen_\sdir = 1$}, as
mentioned in Section~\ref{sec:stretching}.  We do not consider other
relationships in the present work.

Armed with the constraint~\eqref{eq:constr1}, we find that the growth of the
component~$\current_\udir$, given by~\eqref{eq:judir}, is actually
proportional to~$\maxnug$ defined by Eq.~\eqref{eq:maxnug}, and
not~$\geigen_\udir^2$.  This means that the dominant contribution
to~$\current^2$ comes from~\eqref{eq:jmdir},
\begin{align}
\permbfs^2\,\detmetric\current^2 &=
	\permbfs^2\l(\geigen_\udir^2\current_\udir^2
	+ \geigen_\mdir^2\current_\mdir^2
	+ \geigen_\sdir^2\current_\sdir^2\r)\nonumber\\
&= \geigen_\udir^4\geigen_\mdir^2\l(\bb_\udir\,\sdiruv\cdot\l[
		\gradlc\log\l(\geigen_\udir^2|\bb_\udir|\r) -
		(\udiruv\cdot\gradlc)\udiruv\r]\r)^2
	+ \Order{\geigen_\udir^2\,\max(\geigen_\mdir^4,\detmetric^2)},
\end{align}
so that~$\current^2\sim\geigen_\udir^4\geigen_\mdir^2/\detmetric^2$ rather
than the value $\geigen_\udir^6/\detmetric^2$ derived in
Section~\ref{sec:ejtraj}.  With this new scaling the revised dissipation time
scale becomes~\hbox{$\dissts \sim
\detmetric\geigen_\udir^{-2}\geigen_\mdir^{-2}$} instead
of~$\detmetric\geigen_\udir^{-4}$.  Considerably more magnetic energy can thus
be created before resistivity comes into play.  In fact, for the typical case
of~$\geigen_\mdir$ nonexponential (\ie, for steady flows) the time for ideal
evolution is roughly doubled.  Of course because the production of magnetic
energy is exponential in the ideal phase, this leads to an exponentially
larger amount of energy.

The modifications to the growth of the parallel current when the constraint is
taken into account is even more drastic.  By parallel current, we mean the
part of~$\currentv$ along the magnetic field,
\begin{equation}
	\currentp^2 \ldef (\currentv\cdot\Bv)^2/\B^2\,.
	\label{eq:jparadef}
\end{equation}
In Refs.~\cite{Boozer1992,Tang2000} it was concluded that~$\currentv$ aligns
with~$\Bv$ because the dominant part of~$\current^2$ comes
from~$\current_\udir$, so that~\hbox{$\currentp^2 \sim
\geigen_\udir^6/\detmetric^2$}.  But if we apply the
constraint~\eqref{eq:constr1}, we find instead that~\hbox{$\currentp^2 \sim
\geigen_\udir^2\maxnug^2/\detmetric^2$}.  This is a radically different growth
rate: for a steady incompressible flow (\ie, $\geigen_\mdir$ nonexponential
and~$\detmetric=1$), the ratio of the previous to the new result
is~\hbox{$\geigen_\udir^6/\geigen_\udir^2 = \geigen_\udir^4$}.  The most
important implication is that we now have~\hbox{$\currentp^2 \ll
\current_\perp^2$}, so that the current does not align with the magnetic
field.  In terms of length scales, we have that the scale of variation
of~$\Bv$ along itself,~$\ell_\parallel$, is much greater than the
perpendicular variation,~$\ell_\perp$.  The dominant contribution
to~$\currentp$ comes from a mixture of~$\current_\udir$ and~$\current_\mdir$,
whereas the dominant contribution to~$\current^2$ comes from~$\current_\mdir$.

Note that when we speak of alignment of two vectors we are expressing this in
terms of their scalar product.  The scalar product is invariant under
coordinate transformations, including the transformation between Eulerian and
Lagrangian coordinates.  This means that the alignment occurs both in the
Lagrangian and Eulerian frames.  This is the power of the Lagrangian
trajectory approach: the behavior of the vectors is easily derived in
Lagrangian coordinates, and the results automatically apply to the Eulerian
frame if expressed in terms of scalar quantities.

\section{A Singular Initial Condition}
\label{sec:singinit}

So far we have only made use of the constraint~\eqref{eq:constr1}.  In the
present section we offer an application of the other constraints.  If we apply
the constraints~\eqref{eq:constr3} and~\eqref{eq:constr2} respectively to
the~$\current_\udir$ and~$\current_\mdir$ components of the current,
equations~\eqref{eq:judir} and~\eqref{eq:jmdir}, we find
\begin{align}
\permbfs\,\current_\udir &= \geigen_\udir^2\,\bb_\udir
		\,\udiruv\cdot\curllc\udiruv
	- \geigen_\mdir^2\,\bb_\mdir\,\sdiruv\cdot
		\gradlc\log\bigl(\widetilde\geigen_\mdir|\bb_\mdir|\bigr)
	+ \geigen_\udir^{-1}\,
	\Order{\geigen_\mdir\,,1\,,\geigen_\mdir^3/\sqrt{\detmetric}},
	\label{eq:judir2}\\
\permbfs\,\current_\mdir &=
	\geigen_\udir^2\,\bb_\udir\,\sdiruv\cdot
		\gradlc\log\bigl(\widetilde\geigen_\udir|\bb_\udir|\bigr)
	+ \frac{\geigen_\udir}{\geigen_\mdir^2}\,
	\Order{\geigen_\mdir\,,1\,,\geigen_\mdir^3/\sqrt{\detmetric}},
	\label{eq:jmdir2}
\end{align}
where we have made use of the fact that, asymptotically, the Lagrangian
derivatives of~$\log\geigen_\udir$,~$\log\geigen_\mdir$,
and~$\log{\detmetric}$ satisfy~\cite{Thiffeault2002}
\begin{align}
	\sdiruv\cdot\gradlc\log\geigen_\udir &=
		\sdiruv\cdot\gradlc\log\widetilde\geigen_\udir
		+ \Order{\geigen_\sdir\,,\geigen_\sdir/\geigen_\mdir\,,
			\geigen_\mdir/\geigen_\udir},
	\label{eq:sdotgradu}\\
	\sdiruv\cdot\gradlc\log\geigen_\mdir &=
		\sdiruv\cdot\gradlc\log\widetilde\geigen_\mdir
		+ \Order{\geigen_\sdir\,,\geigen_\sdir/\geigen_\mdir\,,
			\geigen_\mdir/\geigen_\udir},
	\label{eq:sdotgradm}\\
	\sdiruv\cdot\gradlc\log{\detmetric} &=
		\sdiruv\cdot\gradlc\log{\widetilde\detmetric}
		+ \Order{\geigen_\sdir\,,\geigen_\sdir/\geigen_\mdir},
	\label{eq:sdotgradg}
\end{align}
with~$\widetilde\geigen_\udir$,~$\widetilde\geigen_\mdir$,
and~$\widetilde\detmetric$ time-independent functions of~$\lav$, unique up to
a multiplicative constant.
Equations~\eqref{eq:sdotgradu}--\eqref{eq:sdotgradg} reflect the contraction
of coordinates along the~$\sdiruv$ direction, which leads to exponential
convergence of Lagrangian derivatives; these convergence rates were derived in
detail in Ref.~\cite{Thiffeault2002}.

From~\eqref{eq:jmdir2}, we see that there are two special initial conditions
for~$\bb_\udir$.  The first is to simply let~\hbox{$\bb_\udir\equiv0$}, but
this is not relevant here because then the magnetic energy does not grow at
the rate~$\geigen_\udir$.  A more interesting possibility is to let
\begin{equation}
	\bbv\cdot\udiruv = c\,\,\widetilde\detmetric^{1/2}\,
		\widetilde\geigen_\udir^{-1},
	\label{eq:singinit}
\end{equation}
where~$c$ is a constant and~$\widetilde\detmetric$
and~$\widetilde\geigen_\udir$ are defined in~\eqref{eq:sdotgradu}
and~\eqref{eq:sdotgradg}.  The~$\widetilde\detmetric^{1/2}$ factor appears
in~\eqref{eq:singinit} because of the definition~$\bb_\udir \ldef
(\udiruv\cdot\bbv/\sqrt{\detmetric})$.  Then the first term
in~\eqref{eq:jmdir2} vanishes, and we find that the leading-order behavior of
the current is
\begin{equation}
	\current^2 \sim \geigen_\udir^2\,\maxnugg/\detmetric^2,
		\qquad
		\maxnugg \ldef
		\max(\maxnug^2\,,\detmetric\,\geigen_\mdir^{-2}),
	\label{eq:jasymsing}
\end{equation}
in contrast to~$\geigen_\udir^{4}\geigen_\mdir^{2}/\detmetric^2$ for a generic
initial condition.  For an asymptotic rate such as~\eqref{eq:jasymsing}, the
dissipation time scale~\eqref{eq:disstsdef} is~\hbox{$\dissts \sim
\detmetric/\maxnugg$}; for a steady incompressible flow (\ie, $\geigen_\mdir$
nonexponential and~$\detmetric=1$), we have~\hbox{$\dissts \sim 1$}.  Hence,
for this special initial condition the dissipation time can be of order unity.
We thus have an exponentially growing magnetic field whose growth can be
sustained for a relatively long time before dissipation becomes important.

As for the parallel current~$\currentp^2$, it remains unchanged for the
singular initial condition and is equal
to~$\geigen_\udir^2\,\maxnug^2/\detmetric^2$; its main contribution is
from~$\current_\udir$.  Thus, asymptotically, the singular initial condition
has the magnetic field and current aligned.  (In terms of the scale of
variation of the magnetic field, we have~\hbox{$\ell_\parallel \sim
\ell_\perp$}.)

The existence of a singular initial condition is made possible by the
constraints~\eqref{eq:constr1} and~\eqref{eq:constr2}--\eqref{eq:constr3}, but
it is also crucial that the Lagrangian derivatives~\eqref{eq:sdotgradu}
and~\eqref{eq:sdotgradg} converge to well-defined time-asymptotic values,
which is the case for almost all initial conditions~\cite{Thiffeault2002}.
Otherwise it would not be feasible to ask for an initial condition of the
form~\eqref{eq:singinit} since~$\bbv$ is constant in time (until the diffusive
regime is reached).

We shall comment further on the singular initial condition in the Discussion
section at the end of the paper.

\section{Helicity Generation}
\label{sec:helicity}

In the present section we turn our attention to the production of magnetic
helicity.  The magnetic helicity $\helic \ldef \Av\cdot\Bv$, where $\Av$ is
the magnetic potential with $\curl\Av = \Bv$, evolves according to
\begin{equation}
	\frac{\pd\helic}{\pd\time} + \div(\vel\,\helic)
	= -\resist\,(2\,\currentv\cdot\Bv - \div(\Av\times\currentv)).
	\label{eq:helicevol}
\end{equation}
(Some terms were absorbed by an appropriate choice of gauge.)  Helicity can
only be created or destroyed if $\resist\ne 0$.  (We refer to the terms on the
right-hand side of~\eqref{eq:helicevol} as the helicity generation terms, but
they are not sign-definite and can also destroy helicity.)  The second
helicity generation term in~\eqref{eq:helicevol} creates no net helicity, as
it vanishes upon integration over the entire volume (for suitable boundary
conditions).  However, it can move helicity from one point to another: it
represents a flux of helicity, so that the total helicity flux
is~\hbox{$\vel\,\helic - \resist\,\Av\times\currentv$}.

The magnetic energy $\EnerB = \B^2/2\permbfs$ is dissipated by Ohmic heating
in the plasma at a rate given by~$\resist\,\current^2$.  We want to examine
the relative magnitude along fluid trajectories of local helicity generation
as given by~\eqref{eq:helicevol} and energy dissipation through Ohmic
heating~$\resist\,\current^2$.  This will help determine whether the net
helicity in a region due to relative transport (helicity flux) or creation
(the~$2\eta\,\currentv\cdot\Bv$ term).  It is thought that helicity generation
is an important ingredient for the creation of a large-scale magnetic
field~\cite{Boozer1993}, in part because helicity decays on a slower timescale
than energy~\cite{Taylor1974} so its presence ensures a long-lived magnetic
field.  It is also important because of its implications for the topology
(knottedness) of magnetic field
lines~\cite{Woltjer1958,Moffatt1969,Moffatt1985}.  Given the new results
involving the constraints, is reasonable to ask if there is sufficient time in
the ideal energy evolution phase to create a significant amount of helicity.
Put another way, when the Ohmic heating term becomes of order one, how large
are the helicity generation terms?

Following the trajectory of a fluid element, \eqref{eq:helicevol} becomes
\begin{equation}
	\l.\frac{\pd}{\pd\time}\r|_{\lav}(\Av\cdot\bbv)
	= -\resist\,(2\,\currentv\cdot\bbv
		- \divlc(\Av\times\currentv))
	\label{eq:helicevollagr}
\end{equation}
where all the variables are expressed in the Lagrangian frame
and~\hbox{$\helic = \Av\cdot\bbv/\sqrt{\detmetric}$}.  If we form the ratio of
the helicity generation terms on the right-hand side
of~\eqref{eq:helicevollagr} and power dissipation~$\resist\,\current^2$,
the resistivity~$\resist$ cancels, making these two ratios convenient measures
of the relative strengths of the effects.  We can now ascertain whether the
helicity generation terms become important before ideal evolution ends, and
find which helicity generation term is largest.

We start by comparing the first term on the right-hand side
of~\eqref{eq:helicevollagr} to the Ohmic heating term~$\resist\,\current^2$,
ignoring the constraint~\eqref{eq:constr1}:
\begin{equation}
	\frac{\currentv\cdot\bbv}{\current^2} =
		{\sqrt{\detmetric}}\,\frac{\currentp\,\B}{\current^2}
		\sim {\sqrt{\detmetric}}\,
			\frac{\geigen_\udir^3}{\detmetric}\,
			\frac{\geigen_\udir}{\sqrt{\detmetric}}
			\frac{\detmetric^2}{\geigen_\udir^6}
		= \detmetric\,\geigen_\udir^{-2}\,.
	\label{eq:jdotbnoconstr}
\end{equation}
The helicity generation term is typically exponentially small compared to
energy dissipation.  Including the constraint~\eqref{eq:constr1} does not
improve matters:
\begin{equation}
	\frac{\currentv\cdot\bbv}{\current^2}
		\sim {\sqrt{\detmetric}}\,
		\frac{\geigen_\udir\,\maxnug}{\detmetric}\,
			\frac{\geigen_\udir}{\sqrt{\detmetric}}\,
			\frac{\detmetric^2}{\geigen_\udir^{4}\geigen_\mdir^{2}}
		= \detmetric\,\geigen_\udir^{-2}\,\geigen_\mdir^{-2}\,\maxnug,
	\label{eq:jdotbconstr}
\end{equation}
so that the helicity generation term is still typically exponentially smaller
than Ohmic heating.

For the singular initial condition of Section~\ref{sec:singinit}, the ratio of
helicity generation to Ohmic dissipation yields
\begin{equation}
	\frac{\currentv\cdot\bbv}{\current^2}
		\sim {\sqrt{\detmetric}}\,
		\frac{\geigen_\udir\,\maxnug}{\detmetric}\,
			\frac{\geigen_\udir}{\sqrt{\detmetric}}\,
			\frac{\detmetric^2}
			     {\geigen_\udir^2\,\maxnugg}
		= \frac{\detmetric\,\maxnug}{\maxnugg};
	\label{eq:jdotbsing}
\end{equation}
The ratio is nonexponential (\ie, order one) for a steady incompressible flow.
This opens the possibility of creating a sizable amount of helicity before
Ohmic dissipation becomes important.

The second term on the right of~\eqref{eq:helicevollagr}---the flux of
magnetic helicity in Lagrangian coordinates---is given by
\begin{equation}
   \divlc(\Av\times\currentv) = \sum_{\mu=1}^3
   \divlc\l[
     \geigen_\mu^2\,(\Av\times\ediruv_\mu)\,\current_\mu/{\sqrt\detmetric}
     \r]
   \label{eq:Adotcurljexp}
\end{equation}
where~\hbox{$\mu = \{\udir,\mdir,\sdir\}$} and~\hbox{$\{\ediruv\} \ldef
\{\udiruv,\mdiruv,\sdiruv\}$}.  Because~$\Av$ does not acquire a special
orientation with respect to~$\ediruv_\mu$, upon expansion the divergence
in~\eqref{eq:Adotcurljexp} will generate terms of the
form~\hbox{$\udiruv\cdot\gradlc\log\geigen$}, and these grow proportionally
to~$\geigen_\udir$~\cite{Thiffeault2002}.  Hence we find
\begin{equation}
\divlc(\Av\times\currentv) \sim \frac{\geigen_\udir}
		{\sqrt{\detmetric}}
	\sum_{\mu=1}^3 \geigen_\mu^2\,\current_\mu\,.
	\label{eq:Adotcurljexp2}
\end{equation}
The overall growth rate thus depends on whether the
constraint~\eqref{eq:constr1} is used, because this affects which component
of~$\current_\mu$ is largest.  Without the constraint we have
both~$\current_\udir$ and~$\current_\mdir$ proportional
to~$\geigen_\udir^2/\sqrt{\detmetric}$; then the dominant contribution
in~\eqref{eq:Adotcurljexp2} comes from~$\current_\udir$, yielding
\begin{equation}
\divlc(\Av\times\currentv) \sim
	\frac{\geigen_\udir}{\sqrt\detmetric}\,
	\geigen_\udir^2\,\current_\udir
	\sim {\geigen_\udir^5}/{\detmetric}\,.
	\label{eq:Adotcurljexpnoconstr}
\end{equation}
But the result~\eqref{eq:Adotcurljexpnoconstr} was obtained without taking
differential constraints into account and is thus incorrect.  Upon applying
the constraint~\eqref{eq:constr1}, we have~\hbox{$\current_\udir \sim
\maxnug/\sqrt{\detmetric}$}, so that
\begin{equation}
\divlc(\Av\times\currentv) \sim
      {\geigen_\udir^3}\,\maxnug/{\detmetric}.
      \label{eq:Adotcurljexpconstr}
\end{equation}
Taking the ratio with the Ohmic heating~$\resist\,\current^2$, we find
that~$\divlc(\Av\times\currentv)/\current^2$ is proportional
to~\hbox{$\detmetric\,\geigen_\udir^{-1}$} without the
constraint~\eqref{eq:constr1}, and
to~\hbox{$\detmetric\geigen_\udir^{-1}\geigen_\mdir^{-2}\,\maxnug$} with the
constraint.  For the case of a steady incompressible flow, this helicity
source term is again exponentially smaller than the power dissipation, and
moreover the ratio of the two is roughly~$\geigen_\udir^{-1}$ both with and
without the constraint.  The rate~\eqref{eq:Adotcurljexpconstr} also applies
to the singular initial condition, but in that case~$\current^2$ grows more
slowly [Eq.~\eqref{eq:jasymsing}], allowing more accumulation of helicity
before the end of ideal evolution.

If we compare the two terms (helicity creation and helicity flux) on the
right-hand side of~\eqref{eq:helicevollagr} to each other, we find
\begin{equation}
\frac{\divlc(\Av\times\currentv)}{\currentv\cdot\bbv}
	\sim \geigen_\udir
\end{equation}
in all three cases (constrained, unconstrained, and singular initial
condition).  We conclude that though the proper application of the constraints
changes the absolute magnitude of each helicity source term
in~\eqref{eq:helicevollagr}, when expressed as a ratio with the current their
relative magnitude is unchanged.  The helicity flux (second term) always tends
to dominate the helicity creation (first term) as a helicity source on the
right-hand side of~\eqref{eq:helicevollagr}.

Table~\ref{tab:constrcomp} summarizes the growth rates derived in previous
sections and in this section for helicity generation.  For a clearer picture
of relative magnitudes, Table~\ref{tab:constrincomp} gives the corresponding
growth rates for a steady incompressible flow.  The singular initial condition
is the only case that allows for the possibility of significant helicity
creation or flow before the dissipative regime is reached.  For the other two
cases (unconstrained and constrained), for very small~$\resist$ the magnetic
field will have built up huge gradients by the time the evolution ceases to be
ideal.  Exponentially large amounts of power thus seem required to create or
move helicity.  Even for the singular initial condition, there is no guarantee
that the helicity will be created rather than destroyed, or will be
concentrated rather than dispersed, since the generation and flow terms are not
definite in sign.  In fact, it is known that substantial cancellation of the
helicity occurs~\cite{Gilbert2002}, a fact that is not captured in our
analysis.

We emphasize that the results for the unconstrained case are presented merely
for comparison purposes: the constrained results always apply because
differential constraints are an unavoidable geometrical consequence of the
chaotic nature of the flow~\cite{Thiffeault2002}.  The singular initial
condition, however, only applies to situations that are initialized in the
special manner given by~\eqref{eq:singinit}.  Otherwise the constrained result
applies.

\begin{table*}
\begin{tabular}{cccccl}
& \textsl{No constraint} & \hspace*{.5em} & \textsl{Constraint}
  & \hspace*{1em} \textsl{Sing. initial cond.} \hspace*{1em}
  & \textsl{Comments} \\
\hline
$\B^2$ & $\geigen_\udir^2/\detmetric$ &&
	$\geigen_\udir^2/\detmetric$ &
	$\geigen_\udir^2/\detmetric$ & Magnetic energy\\
$\current^2$ & $\geigen_\udir^{6}/\detmetric^2$ &&
	$\geigen_\udir^{4}\geigen_\mdir^{2}/\detmetric^2$ &
	$\geigen_\udir^2\,\maxnugg/\detmetric^2$ & Power dissipation\\
$\currentp^2$ & $\geigen_\udir^6/\detmetric^2$ &&
	$\geigen_\udir^2\,\maxnug^2/\detmetric^2$ &
	$\geigen_\udir^2\,\maxnug^2/\detmetric^2$ & Parallel current\\
$\currentv\cdot\bbv/\current^2$ &
	$\detmetric\geigen_\udir^{-2}$ &&
	$\detmetric\geigen_\udir^{-2}\,\geigen_\mdir^{-2}\,\maxnug$ &
	$\detmetric\,\maxnug/\maxnugg$ & Helicity creation
	vs power\\
$\divlc(\Av\times\currentv)/\current^2$ &
	$\detmetric\geigen_\udir^{-1}$ &&
	$\detmetric\geigen_\udir^{-1}\,\geigen_\mdir^{-2}\,\maxnug$ &
	$\detmetric\geigen_\udir\maxnug/\maxnugg$
	& Helicity flux vs power\\
\hline
\end{tabular}
\caption{Asymptotic growth of physical quantities.  Three cases are
considered: (i) No differential constraint; (ii) Constraint given by
Eq.~\eqref{eq:constr1} used; (iii) Singular initial condition of
Section~\ref{sec:singinit}, which uses the
constraints~\eqref{eq:constr1}--\eqref{eq:constr3}.  The factor~$\maxnug$ is
defined in~\eqref{eq:maxnug},~$\maxnugg$ in~\eqref{eq:jasymsing}.}
\label{tab:constrcomp}
\end{table*}

\begin{table}
\begin{tabular}{ccccc}
& \textsl{No constraint} & \hspace*{.5em} & \textsl{Constraint}
  & \hspace*{.75em} \textsl{Sing. initial cond.} \hspace*{.75em}\\
\hline
$\B^2$ & $\geigen_\udir^2$ &&
	$\geigen_\udir^2$ &
	$\geigen_\udir^2$ \\
$\current^2$ & $\geigen_\udir^{6}$ &&
	$\geigen_\udir^{4}$ &
	$\geigen_\udir^2$ \\
$\currentp^2$ & $\geigen_\udir^6$ &&
	$\geigen_\udir^2$ &
	$\geigen_\udir^2$ \\
$\currentv\cdot\bbv/\current^2$ &
	$\geigen_\udir^{-2}$ &&
	$\geigen_\udir^{-2}$ & $1$ \\
$\divlc(\Av\times\currentv)/\current^2$ &
	$\geigen_\udir^{-1}$ &&
	$\geigen_\udir^{-1}$ &
	$\geigen_\udir$ \\
\hline
\end{tabular}
\caption{As for Table~\ref{tab:constrcomp} but assuming a steady
incompressible flow ($\geigen_\mdir\sim 1$, $\detmetric=1$).}
\label{tab:constrincomp}
\end{table}

\section{Discussion}
\label{sec:discussion}

Using Lagrangian coordinates, we have revised earlier estimates of the
magnitude of energy dissipation (Section~\ref{sec:constr}) and helicity
generation (Section~\ref{sec:helicity}).  The estimates were corrected for
differential constraints on the rates and directions of stretching.  The
inclusion of constraints leads roughly to a doubling of the period of ideal
evolution, for the case of a steady incompressible flow.  We also found after
using constraints that he current does not align with the magnetic field in
the ideal phase, in contrast with the results of
Refs.~\cite{Boozer1992,Tang2000} where it was concluded that it did.  The
constraints thus have a deep implication on the evolution of the magnetic
field, at least in the ideal phase.

In spite of the constraints, we find that the helicity generation terms are
exponentially smaller than the energy dissipation, so that creation of
significant amounts of helicity cannot occur in the period of ideal energy
growth (when resistivity can be neglected).  Our analysis does not take into
account the fact that the helicity created will be fractal in nature, and so
very little of it will contribute to a large-scale magnetic field, as pointed
out by Gilbert~\cite{Gilbert2002}.  This suggests a close inspection of the
energy dissipation in numerical simulations, since current dynamo models could
require vast amounts of energy to create helicity.

We found (Section~\ref{sec:singinit}) that there exists a special initial
condition with desirable properties: (i) the power dissipated is of the same
order as the magnetic energy; (ii) the helicity generation terms are
comparable to the power dissipated.  This opens the possibility of a flow
creating a considerable amount of energy and helicity before reaching the end
of the ideal evolution regime.  The physical relevance of the singular initial
condition is uncertain.  We cannot expect a realistic system to have an
initial magnetic field corresponding that particular condition.  The part of
the initial condition that is not singular will reach the dissipation scale
much sooner, so the ideal evolution hypothesis becomes invalid.  But does the
singular part of the initial condition survive, or is it dissipated away?  An
analysis involving the full, dissipative equation is necessary to establish
this and has not yet been performed.  If it does survive, it could be closely
related to the ``strange eigenfunction'' (\ie, a fractal) that occurs in the
freely-decaying advection-diffusion problem~\cite{Pierrehumbert1994} and its
analogue for the dynamo~\cite{STF}.

\begin{acknowledgments}
This work was supported by the National Science Foundation and the Department
of Energy under a Partnership in Basic Plasma Science grant,
No.~DE-FG02-97ER54441.
\end{acknowledgments}


\end{document}